\documentclass[12pt]{article}
\usepackage{graphicx}
\def\lsim{\mathrel{\rlap{\lower4pt\hbox{\hskip1pt$\sim$}}
    \raise1pt\hbox{$<$}}}         
\def\gsim{\mathrel{\rlap{\lower4pt\hbox{\hskip1pt$\sim$}}
    \raise1pt\hbox{$>$}}}         

\begin{document}
\begin{titlepage}
\bigskip
\rightline {LYCEN 2000-44}
\bigskip

\begin{center}

{\large {\bf Elastic $pp$ and $\bar pp$ scattering in the \\
               Modified Additive Quark Model}}

\vskip 2.cm

{\bf P. Desgrolard}$^{a,}${\footnote{E-mail: desgrolard@ipnl.in2p3.fr}},
{\bf M. Giffon}$^{a,}${\footnote{E-mail: giffon@ipnl.in2p3.fr}}, {\bf E.
Martynov}$^{b,}${\footnote{E-mail: martynov@bitp.kiev.ua}}.

\end{center}

\bigskip
\bigskip
\bigskip
\noindent $^{a}${\it Institut de Physique Nucl\'eaire de Lyon,
IN2P3-CNRS et Universit\'{e} Claude Bernard,\\ 43 boulevard du 11
novembre 1918, F-69622 Villeurbanne Cedex, France\\}

\noindent $^b${\it Bogoliubov Institute for Theoretical Physics,
National Academy of Sciences of Ukraine,\\ 03143, Kiev-143,
Metrologicheskaja 14b, Ukraine\\}

\bigskip

\bigskip
\bigskip

\begin{minipage}[t]{14.0cm}
{\large\noindent{\bf Abstract}} The modified additive quark model,
proposed recently, allows to improve agreement of the standard additive
quark model with the data on the $pp,\ \bar pp,\ \pi^{\pm} p,\ \gamma p$
and $\gamma \gamma$ total cross-sections, as well as on the ratios of real
to imaginary part of $pp$ and $\bar pp$ amplitudes at $t=0$. Here, we
extend this model to non forward elastic scattering of protons and
antiprotons. A high quality reproduction of angular distributions at 19.4
GeV $\leq \sqrt{s}\leq $ 1800 Gev is obtained. A zero at small $|t|$ in
the real part of even amplitude in accordance with a recently proved high
energy general theorem is found.
\end{minipage}
\end{titlepage}
\section{Introduction}
Small angle elastic scattering of hadrons has always been a crucial
source of information for the dynamics of strong interaction. As a
rule, these processes, being outside the realm of applications of the
theory of the strong interactions (QCD), are described in
approaches based on the $S$-matrix theory. In particular, the various
Regge models are very successful in this direction. However, in the
framework of a Regge approach it is impossible to calculate all the
ingredients needed in the amplitudes and some additional arguments
are used to construct main objects such as Regge trajectories and
residue functions. Sometimes they are derived from the fundamental
theory, but usually they are based on an intuitive physical picture
and on the analysis of the available experimental data.

The additive quark model (AQM)~\cite{aqm} is an example of such a
line of arguments. An amplitude of composite particle interaction is
constructed as a sum of the elementary amplitudes of the interaction
among the constituent quarks. This leads to remarkable relations
(counting rules) between various hadronic cross-sections in rather good
agreement with the experimental data.

In a recent paper~\cite{maqm}, the standard AQM (which we abbreviate in
this old version as SAQM) and the ensuing counting rules are modified to
take into account not only the quark-gluonic content of the Pomeron but
also of the secondary Reggeons, as well as the fact that the soft Pomeron
is not just a gluonic ladder. The new model, which we call Modified
Additive Quark Model (MAQM), is successfully applied to describe various
total cross-sections (nucleon-nucleon, $\pi$ meson-nucleon, $\gamma-p$ and
$\gamma-\gamma$) and the $\rho-$ratio of the real to the imaginary part of
the forward $pp,\, p\bar p$ scattering amplitudes. The next step~: to
consider the differential cross-sections is the object of this paper, {\it
i.e.} we extend and further test the model for $t\neq 0$. Let remind that
there are two couplings of Pomeron with quarks in MAQM: the first one
corresponds to the ordinary vertex quark-Pomeron-quark, another one is new
and describes a "simultaneous" interaction of Pomeron with two quarks in
hadron. It was found from the fit to cross-sections in \cite{maqm} that
the corresponding term in amplitudes is negative. At some $t\neq 0$ an
interference of the ordinary and new term can produce a zero in the
elastic amplitude and consequently a dip structure in the differential
cross-section. This argument was one of reasons to extend MAQM for $t\neq
0$. We deal only with $pp$ and $p\bar p$ elastic scattering because, for
these processes, are available the richest and most precise experimental
data in a wide region of energy $\sqrt{s}$ and momentum transfer $t$. Many
models (for instance~\cite{BornSPs,dgp,cov,GaurNic,DipPom,brazil}) of $pp$
and $p\bar p$ elastic scattering amplitudes describe quite well the
available data (see also the reviews~\cite{reviews} and references
therein). A supercritical Pomeron with the intercept $\alpha_{\cal
P}(0)>1$ is used at the Born level for most of them
\cite{BornSPs,dgp,cov}. Such a Pomeron must then be unitarized because it
does not satisfy the unitarity constraints. Usually a method of
eikonalization~\cite{dgmp} is used to this aim. A method for reproducing
$t\ne 0$ data, based on the model of stochastic vacuum, (attractive by its
success) consists in parametrizing each angular distribution in the $s,t$
space (see for example~\cite{brazil}), the energy dependence of the
amplitude being absorbed in the parameters. The price to pay is of course
the multiplication of the number of free parameters. An alternative way is
to construct a model that from the beginning does not violate the
requirements for the analyticity and unitarity of the scattering amplitude
\footnote{We mean only that the model should not violate grossly and
explicitly the constraints of unitarity. This, unfortunately, does not
guarantee that unitarity is satisfied.}. An example of such a model is the
model of the so called maximal Pomeron and maximal Odderon~\cite{GaurNic}.
However, serious arguments against the maximal Odderon
exist~\cite{dgmp,antimOd}. In~\cite{maqm}, we stick to the last kind of
approach and we use an extended AQM which can be applied not only to
nucleon-nucleon scattering. As an explicit choice, for the Pomeron
contribution we choose a simple model, namely a special case of the soft
Dipole Pomeron, with a unit intercept~\cite{DipPom,dlm2}
 which leads to a high quality
description of the experimental data, both at $t=0$
\cite{dglm,dgm,BlKaWh,cud}
as well as at $t\neq 0$ (we  show it in Sect. 4).

In the present paper we will not consider any eikonalization and work
with the Born amplitudes since in the MAQM they do not violate unitarity
(although with the restricted sense set above).
However,
it is not obvious that eikonalization should not be carried out since
it corresponds to take into account the physical processes of
rescattering corrections. We will not discuss this point here.
The present work may be used as a guide for further investigations.
It should be stressed that the task of reproducing well the entire
set of high energy data (at all values of the momentum transfer $t$),
though far from simple, as a long (and direct) experience teaches
us may seem to have a poor theoretical content~\cite{reviews}.
This is indeed the truth in the sense that we have not yet any means of
determining the soft amplitudes from first principles. However,
we believe that it is important to explore all the
approaches yielding a good agreement with the existing data. To the
extent that they may lead to different extrapolations which,
hopefully and foreseeably, will be checked in future experiments, we
shall have {\it a posteriori} the means of establishing a hierarchy
among them.

In Section 2, we recall the main assumptions
of~\cite{maqm}, focalizing on a few arguments in favor of the chosen
Pomeron used in the MAQM at $t=0$. In Section 3, we formulate our MAQM
extension
for $t\neq 0$. The results of the fit of the MAQM to the experimental
data are
presented and discussed in Section 4. We examine also if the
amplitude that fits very well the data exhibits automatically a zero
in the real part of its even component as required by a general
theorem due to A.~Martin~\cite{martin}. Some items are also discussed
in this Section (concerning in particular the Odderon and the
logarithmic trajectories).
\section{The modified additive quark model at $t=0$}
Let us review the main properties of the MAQM, formulated for the
forward scattering amplitudes (for details see~\cite{maqm}).
\subsection{Pomeron}
The Pomeron contribution to the $pp $ and $\bar pp$
 scattering amplitude at $t=0$ is written  as
\footnote{The Pomeron contributions to the $\pi p,\,
\gamma p$ and $\gamma \gamma$ amplitudes are given in~\cite{maqm}.}
\begin{equation}\label{1}
 A^{(pp)}_{\cal P}(s,0)=  9P_p^2\Big[A^{(1)}_{\cal
P}(s/9,0)+2A^{(2)}_{\cal P}
  (2s/9,0) +  A^{(3)}_{\cal P}(4s/9,0)\Big]\ .
\end{equation}
The choice of the elementary Pomeron quark-quark amplitudes
$A^{(i=1-3)}_{\cal P}$ is very important from a phenomenological point of
view. It is known from a comparison of various Pomeron
models~\cite{dglm,cud} that equivalent good fits to the $t=0$ data on the
$\rho-$ratios of the real to imaginary parts of the forward amplitude
$\rho =\Re eA(s,0)/\Im mA(s,0)$ and on the total elastic cross sections
$\sigma_{tot}$ for meson-nucleon and nucleon-nucleon interactions are
achieved, at $s\to \infty$, either if $\sigma_{tot}\propto \ell n(s/s_0)$,
or $\propto \ell n^2(s/s_0)$ or $\propto (s/s_0)^\epsilon$. Unfortunately,
as far as only $pp$ and $\bar pp$ are concerned, existing data do not
allow to discriminate unambiguously~\cite{dgm} between these three
behaviors. Nevertheless, as was noted in \cite{cud}, the model with
$\sigma_{tot}\propto \ell n(s/s_0)$ is the most "stable" in the sense that
the fitted parameters and $\chi^{2}/d.o.f.$ do not changed in practice
under variation of $s_{min}$ in the energy range 5 - 10 GeV (the models
were fitted to the data at $s\geq s_{min}$). Hence, among the possible
Pomerons, we select a "Dipole Pomeron" with an intercept equal to one,
$\alpha_{\cal P}(0)=1$, {\it i.e.} corresponding to a double pole of the
amplitude in the complex angular momenta plane $j$ and yieding an
asymptotic behavior $\sigma_{tot}\propto \ell n(s/s_0)$ with an economy of
parameters~\cite{cud}. It is interesting to note that such a Dipole
Pomeron is a singularity of the partial amplitude, $\phi (j,t)\propto
(j-\alpha_{\cal P}(t))^{-\gamma}$, with the maximal hardness that does not
violate the evident inequality $\sigma_{elastic}(s)\leq \sigma_{tot}(s)$.
The inequality $\gamma \leq 2$ must be satisfied if the Pomeron trajectory
at small $t$ is linear, $\alpha_{\cal P}(t)\approx 1+\alpha'_{\cal P}t$,
and $\gamma =2$ corresponds to the Dipole Pomeron. The contribution of the
Dipole Pomeron to the forward hadron-hadron elastic scattering amplitude
is written
$$
 A_{\cal P}^{hh}(s,0)=C_1+C_2\ell n(-is/s_0)\ ,
$$
where $C_1$, as it follows from the fit, is a negative constant (we
take consistently $s_0=1$ ${\rm GeV}^2$). This may be surprising
because at small energies the contribution of Dipole Pomeron
to $\sigma_{tot}$ would be negative \footnote{It is noted also
in~\cite{cud}}. However this term can be treated~\cite{dlm}
as an effective contribution of the Pomeron rescatterings and it is
straightforward to demonstrate that its sign may be negative. The
above arguments and those given in~\cite{maqm} suggest that we
may choose to write the Pomeron amplitudes in MAQM at $t=0$ as follows
\begin{equation}\label{2}
\begin{array}{lll}
  A^{(1)}_{\cal P}(s,0)= & ig_1^2\Big[-\zeta_{\cal P}+L(s)\Big]\ , \\
  A^{(2)}_{\cal P}(s,0)= & ig_1g_2\Big[-\zeta_{\cal P}+L(s)\Big]\ ,\\
  A^{(3)}_{\cal P}(s,0)= & ig_2^2\Big[-\zeta_{\cal P}+L(s)\Big]\ ,
\end{array}
\end{equation}
where
$$
L(s)=\ell n(-is/s_0).
$$
\subsection{Secondary Reggeons}
In $pp$ and $\bar p p$ scattering, the secondary Reggeons are numerous,
however,
for the energy range involved here we can choose to keep only $f-$and
$\omega-$Reggeons, two non degenerate C=+1 and C=-1 meson trajectories.
As it is argued in \cite{maqm} instead of nine
identical diagrams for the $f-$Reggeon in the SAQM, leading to the
factor 9, one obtains
\begin{equation}\label{3}
A^{(pp)}_{f}(s,0)=P_p^2(5+4\lambda_f)A^{(qq)}_f(s/9,0),
\end{equation}
where
\begin{equation}\label{4}
  A^{(qq)}_f(s,0)= ig_f^2\bigg (-i{s/s_0} \bigg )^{\alpha_f(0)-1}
\end{equation}
and $\lambda_f$ is a constant taking into account a mixing of $u\bar
u$ and $d\bar d$ quark states in $f-$Reggeon.
Similarly, for $\omega-$Reggeon we set
\begin{equation}\label{5}
A^{(pp)}_{\omega}(s,0) =
P_p^2(5+4\lambda_{\omega})A^{(qq)}_{\omega}(s/9,0)\ ,
\end{equation}
\begin{equation}\label{6}
  A^{(qq)}_{\omega}(s,0)= g_{\omega}^2
  \bigg (-i{s/s_0} \bigg )^{\alpha_{\omega}(0)-1}\ .
\end{equation}
\medskip
An important property of the Dipole Pomeron model is that all fits
give a high value of the $f-$Reggeon intercept, $\alpha_f(0)\approx
0.80\div 0.82$~\cite{dglm,dgm,BlKaWh,cud}. Does such an intercept
contradict the data on the $f-$trajectory known from the resonance
region? The answer is "yes" if the trajectory is assumed to be
linear. However, aside from general theoretical arguments against
linear trajectories, the experimental data on the resonances lying on
the $f-$trajectory indicate its nonlinearity (see Fig.1).
\begin{figure}[ht]\label{fig.1}
\begin{minipage}[t]{7.0cm}
\begin{center}
\includegraphics*[scale=0.38]{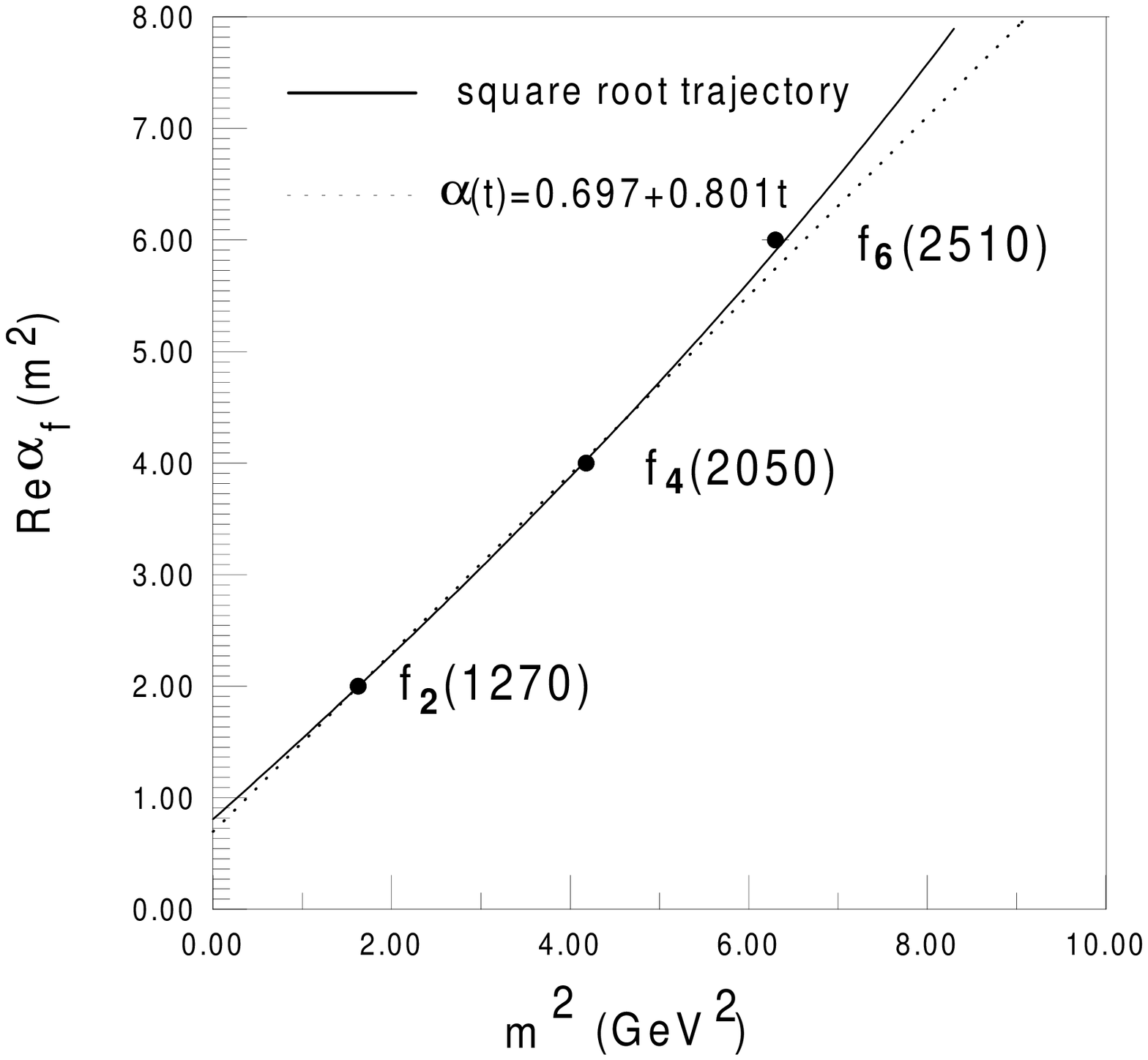}
\caption{Real part of the $f-$trajectory versus $m^2$, squared mass of the
resonance. Solid line is the square root trajectory used in the present
work (see the text). Dashed straight line is the result of a linear fit.}
\end{center}
\end{minipage}
\hskip .9cm
\begin{minipage}[t]{7.0cm}
\begin{center}
\includegraphics*[scale=0.44]{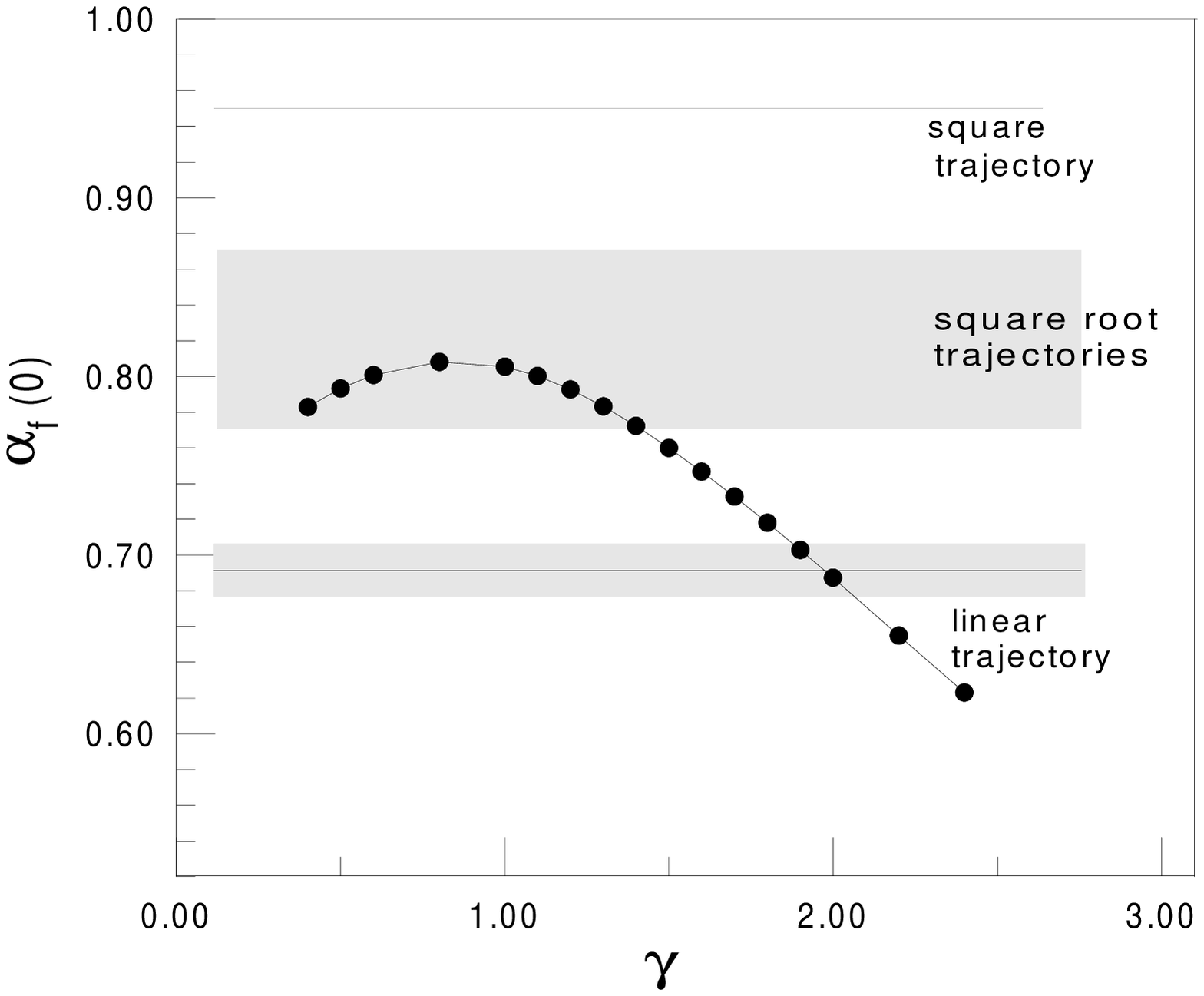}
\vskip 0.2cm
\caption{Intercept of the $f-$trajectory correlated with the
power $\gamma$ of $\ell n s$ in $\sigma_{tot}(s)$. Intervals for an
intercept of various parametrizations are shown.}
\end{center}
\end{minipage}
\end{figure}
The following parabola passes exactly through the three known resonances~:
$\alpha_f(t)=\alpha_f(0)+\alpha'_f t +\beta_ft^2 $ with
$\alpha_f(0)=0.96$, $\alpha'_f=0.59$ GeV$^{-2}$, $\beta_f=0.03$
GeV$^{-4}$. This leads to a too high intercept and can cause problems in
the fit to differential cross-sections at large $|t|$. A more realistic
trajectory such as
$\alpha_f(t)=\alpha_f(0)+\gamma_1(\sqrt{4m_{\pi}^2}-\sqrt{4m_{\pi}^2-t}\
)+ \gamma_2(\sqrt{t_1}-\sqrt{t_1-t}\ ), $ gives $0.77<\alpha_f(0)<0.87$.

Generally, there is an evident correlation between the intercept
of the $f-$Reggeon and the model for the Pomeron. This is due to
the fact that in all known processes, Pomeron and $f-$Reggeon
contribute additively. As a rule, a higher $f-$intercept is
associated with a slower growth with energy due to Pomeron
contribution (as an example see also~\cite{cud}). In Fig.2, we
illustrate this observation and show how $\alpha_f(0)$ is
correlated with a power of $\ell n s$ in the behaviour of the
total cross-section, if the forward scattering amplitudes are
parameterized in the form.
$$
A^{hh}(s,0)=i\Big[C_1+C_2\ell n^{\gamma}(-is/s_0)\Big]+R(s,0),
$$
where the explicit form of secondary Reggeons contribution $R(s,0)$,
depends on the nature of interacting hadrons (see \cite{dglm,cud} for
details).
In our opinion, a good way to fix the intercepts of $f-$ and
$\omega-$Reggeons is after fitting the total cross-sections. Doing so
and avoiding an extra number of parameters, we choose the following form of
$f-$trajectory
$$
\alpha_{f}(t)=\alpha_{f}(0)+\gamma_{f}(\sqrt{t_{f}}-\sqrt{t_{f}-t})\ ,
$$
where the intercept $\alpha_{f}(0)=0.810$ is fixed from the
fit~\cite{maqm} to the total cross-sections, the effective threshold
$t_{f}=14.964 $ GeV$^{2}$ and the parameter $\gamma_f =5.504$ GeV$^{-1}$
are determined from the fit to the positions of the three known
resonances. For $\omega-$trajectory (with only two known resonances and
absence of information on possible higher resonances) we use a linear form
$$
\alpha_{\omega}(t)=\alpha_{\omega}(0)+\alpha'_{\omega}t\ ,
$$
with the intercept $\alpha_{\omega}(0)=0.422$ from \cite{maqm} and with
the slope $\alpha'_{\omega}=0.946$ GeV$^{-2}$ determined in a fit to
resonances.
\section{The modified additive quark model at $t\neq 0$}
The amplitudes under interest are written
\begin{equation}\label{7}
  A^{\bar pp}_{pp}(s,t)=A_{\cal P}(s,t)+A_f(s,t)
  \pm\Big[A_{\cal O}(s,t)+A_{\omega}(s,t)\Big]\ .
\end{equation}
The normalization is
\begin{equation}\label{8}
  \sigma_{tot}(s)=8\pi \Im m A(s,t=0),\qquad
  \frac{d\sigma}{dt}(s,t)=4\pi \left|A(s,t)\right|^2   \ .
\end{equation}
The starting points for a parameterization of all terms in (7) at $t\neq
0$ are the corresponding partial amplitudes defined at $t=0$
in~\cite{maqm} and rewritten in detail in (1-6). A special discussion will
be devoted to the Odderon amplitude $a_{\cal O}(s,t)$ which is out of he
game at $t=0$. Let us recall that, in accordance with the main assumption
of additive quark model (as well as of MAQM \cite{maqm}), there is an
interaction of two (in fact of $3 \times 3$) constituent quarks (or lines
in terms of diagram), each of them carrying only a part of the momentum
$\vec p$. Therefore we must define $s_q=(p_1+p_2)^2=s/9$, for protons
assuming that the whole momentum is distributed equally between all quarks
in each of them. As concerns the $t$ channel, we consider for Pomeron and
$f-$Reggeon a single exchange (one line in $t$ channel in terms of
diagram); our Odderon also is supposed to behave as a one
Reggeon-exchange. Consequently, we have no reason to divide $t$ in the
final amplitude by any number~\footnote{ Such a division, {\it e.g.} by 9,
occurs when three gluon- or three Pomeron- exchanges (three lines in $t$
channel) are considered (as in \cite{DL96}), implying a distribution of
the momentum $\vec q$ along three lines, each of them being assumed to
carry an averaged momentum $\vec q/3$.}.
\subsection{Pomeron }
Starting from (1), the Pomeron contribution at $t\neq 0$ will have the form
\begin{equation}\label{9}
A_{\cal P}(s,t)=9\Big[A_{\cal P}^{(1)}(s/9,t)+2A_{\cal P}^{(2)}(2s/9,t)+
  A_{\cal P}^{(3)}(4s/9,t)\Big]\ .
\end{equation}
The most direct generalization of $A^{(i=1-3)}_{\cal P}(s,0)$ (in (2)) is
to consider all "coupling constants" $g_1, g_2, \zeta_{\cal P}$ as
functions of $t$ and to multiply each $A^{i}_{\cal P}(s,0)$ by the usual
Regge factor $\Bigl (-is/s_0\Bigr )^{\alpha_{\cal P}(t)-1}$. Namely, we
write
\begin{equation}\label{10}
\begin{array}{lll}
  A^{(1)}_{\cal P}(s,t)= &
  ig_1^2(t)\Big[-\zeta_{\cal P}(t)+ L(s)\Big]
  \Bigl (-is/s_0\Bigr )^{\alpha_{\cal P}(t)-1}\ ,\\
  A^{(2)}_{\cal P}(s,t)= &
  ig_1(t)g_2(t)\Big[-\zeta_{\cal P}(t)+ L(s)\Big]
  \Bigl (-is/s_0\Bigr )^{\alpha_{\cal P}(t)-1}\ ,\\
  A^{(3)}_{\cal P}(s,t)= &
  ig_2^2(t)\Big[-\zeta_{\cal P}(t)+ L(s)\Big]
  \Bigl (-is/s_0\Bigr )^{\alpha_{\cal P}(t)-1}\ .
\end {array}
\end{equation}
As the simplest variant we choose the linear Pomeron trajectory
(again with an intercept equal 1)
\begin{equation}\label{11}
  \alpha_P(t)=1+\alpha'_Pt\  ;
\end{equation}
we considere also the case of a logarithmic trajectory which
is discussed in details in Section 4.
Of course, more sophisticated Pomeron models can be proposed but they lead
to an extra number of parameters and we will not consider them.
Finally, to avoid proliferation of parameters, we will assume simple
exponential "residue functions" $g_{i=1,2}$ and
$\zeta_{\cal P}$
\begin{equation}\label{12}
  g_1(t)=g_1\exp(b_1t),\qquad g_2(t)=g_2\exp(b_2t),\qquad
  \zeta_{\cal P}(t)=\zeta_{\cal P}\exp(b_{\zeta_{\cal P}}t).
\end{equation}
Thus there are seven
($g_1, g_2, \zeta_{\cal P}, b_1, b_2, b_{\zeta_{\cal P}},$ $
\alpha'_{\cal P}$) parameters for the Pomeron term of amplitude.
With this Pomeron model the Froissart-Martin bound is not violated
and the total cross-section behaves as $\ell ns$ when $s\to\infty$.
\subsection{Secondary Reggeons}
Generalizing~\cite{maqm}, the $f-$Reggeon amplitude in the MAQM
is written as
\begin{equation}\label{13}
  A_f(s,t)=(5+4\lambda_f)A_f^{(qq)}(s/9,t)\ ,
\end{equation}
where
\begin{equation}\label{14}
  A_f^{(qq)}(s,t)=ig_f^2\Bigl(-i\frac{s}{s_0}\Bigr)^
  {\alpha_f(t)-1}e^{b_ft}, \quad\alpha_f(t)=\alpha_f(0)+
  \gamma_{f}(\sqrt{t_{f}}-\sqrt{t_{f}-t}).
\end{equation}
As already noted above, for that $f-$Reggeon trajectory we choose a
square-root dependence on $t$, which is more suitable than a linear one,
and fix its parameters from the fits to
cross-sections and resonances (see Subsection 2.2).
The value of $\lambda_f$ is unimportant if only $pp$ and $\bar pp$
processes are considered (it is equivalent to redefine the coupling
$g_f$). Nevertheless, for the present work we keep $\lambda_f=0.439$ which
was obtained from the fit at $t=0$ (see~\cite{maqm}). The number of free
parameters for a fixed trajectory is then only two ($g_f, b_f$).
Following the previous
considerations, we write for the $\omega$-Reggeon
\begin{equation}\label{15}
 A_{\omega}(s,t)=
  (5+4\lambda_{\omega})A_{\omega}^{(qq)}(s/9,t)\  ,
\end{equation}
where
\begin{equation}\label{16}
A_{\omega}^{(qq)}(s,t) =g_{\omega}^2\ \Bigl(-i\frac{s}{s_0}\Bigr)^
  {\alpha_{\omega}(t)-1}e^{b_{\omega}t},\quad
 \alpha_{\omega}(t)=\alpha_{\omega}(0)+\alpha'_{\omega}t\ .
\end{equation}
For the $\omega-$Reggeon trajectory, we recall that we choose a linear
dependence on $t$,
with the parameters given in Subsection 2.2.
In ~\cite{dlm2} a multiplicative factor $r_{\omega}(s,t)$ was introduced in
$A_{\omega}^{(qq)}(s,t)$
to describe a so called "cross-over phenomenon", namely, a zero at small
$t$ in the difference of the $\bar pp$ and $pp$ differential
cross-sections. Here, we extend the kinematic region under consideration up
to $|t|\approx 14$ GeV$^2$ and also include Odderon contributions. An
interference of various odd terms in amplitude could produce now the
mentioned zero automatically and thus we set $r_{\omega}(s,t)=1$.
Again, repeating the arguments given above for the $f-$Reggeon, we put
$\lambda_{\omega}=1.$ When the trajectory is fixed, we are left with only
two free parameters ($g_{\omega},b_{\omega}$).
\subsection{Odderon}
This crossing-odd contribution (added to the
$\omega-$Reggeon ) is a quite delicate and ambiguous point, lacking
sufficiently precise and numerous data. A widespread consensus
\cite{dglm,dgm,cud}, however, is that an Odderon contribution while
insignificant at $t=0$ is very relevant in the large $|t|$ domain.

As compared with the previous contributions, it is important
to notice that it is impossible to apply to the Odderon any additive
quark model rule. Odderon, in contrast with Pomeron and secondary
Reggeons, interacts with the whole proton rather than with
separate quarks since three gluons (or the Odderon by Donnachie and
Landshoff~\cite{dl}) couple simultaneously with three quarks in each $p$ or
$\bar p$.

As repeatedly mentioned, in this paper we stick to the Born
approximation and rescattering corrections are not taken into
account. This is known to be inadequate from the conceptual point of
view and not just for practical reasons of restoring
unitarity when  the Born approximation yields
its violation. The point is particularly delicate concerning the
Odderon which, by universal consensus, should be important at
large $|t|$. For this reason, we parameterize the
Odderon, somewhat artificially, as the sum of two contributions which
we denote as "soft" and "hard"
\begin{equation}\label{17}
A_{\cal O}(s,t)=A_{\cal O}^{(s)}(s,t)+A_{\cal O}^{(h)}(s,t) \ .
\end{equation}
As it is known, the contribution of Odderon at $t=0$ is negligible. We
take into account this fact multiplying both components
by a factor vanishing at $t=0$. For the soft Odderon we
assume like for the Pomeron a dipole form, suitably damped~\footnote{
Strictly speaking it has dipole form only if $\mu=0$.}
\begin{equation}\label{18}
A_{\cal O}^{(s)}(s,t)=g_{{\cal O}s}(t) \Bigl[(1-e^{\beta_st})
                   \ell n(-is/s_0) \Bigr]^\mu
  \Bigl[-\zeta_{{\cal O}s}(t)+\ell n(-is/s_0)\Bigr]
  \bigg(-i\frac{s}{s_0}\bigg)^{\alpha_{\cal O}(t)-1},
\end{equation}
while for the hard one, we choose
\begin{equation}\label{19}
  A_{\cal O}^{(h)}(s,t)=g_{{\cal O}h}(1-e^{\beta_ht})[\ell n(-is/s_0)]^\nu
  \frac{1}{(1-t/t_{{\cal O}h})^4}\ .
\end{equation}

We should give here a few comments concerning the choice of the
Odderon amplitude defined by the above equations.

\noindent 1) The contribution of the soft Odderon to $\sigma_{elastic}$ is
dominated by the region where $|t|$ is small. In this domain the factor
$(1-e^{\beta_st}) \ell n(-is/s_0)$ is nearly constant. It means that the
soft Odderon does not violate the evident inequality $\sigma_{elastic}\leq
\sigma_{tot}$ at any value of $\mu$.

\noindent
2) At the same time the amplitude should not have a singularity at
$t=0$, therefore $\mu $ must be an integer. In the fits we have
considered $\mu=1$ and $\mu=2$.

\noindent
3) The hard Odderon does not exponentially decrease
with $|t|$, therefore it does not violate the restriction
$\sigma_{elastic}\leq \sigma_{tot}$ at any $s$ only if $\nu\leq 1/2$.

\noindent
4) We consider a linear Odderon trajectory but
with a non unit intercept, only constrained by unitarity~\footnote{
For a discussion of a possible intercept less than one for the
Odderon, see for example~\cite{dgmp} and references therein.}
\begin{equation}\label{20}
  \alpha_{\cal O}(t)=1+\delta_{\cal O}+\alpha'_{\cal O}t\ ,
  \qquad \delta_{\cal O}\leq 0 \ .
\end{equation}
In fact, as it will be emphasized below, only the upper bound of the
intercept may be kept. \noindent 5) As for the Pomeron, the soft residue
functions are taken in an ordinary exponential form
\begin{equation}\label{21}
  g_{{\cal O}s}(t)=g_{{\cal O}s}e^{{b_{\cal O}}t}, \qquad
  \zeta_{{\cal O}s}(t)=\zeta_{{\cal O}}e^{b_{\zeta_{\cal O}}t}\ .
\end{equation}
The case of a logarithmic trajectory is reserved for discussion.
Thus the Odderon contribution is controlled by a maximum of ten additional
parameters : $g_{{\cal O}s}, \zeta_{\cal O}, \beta_s,$ $b_{\cal O},
b_{\zeta_{\cal O}}, \delta_{\cal O}, \alpha'_{\cal O}$,\- $g_{{\cal O}h},
t_{{\cal O}h}, \beta_h$.

\medskip
The grand total number of free parameters for MAQM is twenty six, however,
this number will be reduced by fixing some of them by virtue of special
arguments. For example, this paper being devoted to $pp$ and $\bar
pp$ angular distributions, all coupling constants and intercepts are
fixed from the fit to $\sigma_{tot}$ and $\rho$ at $t=0$ as
reported in~\cite{maqm}; furthermore, the remaining parameters of Reggeons
trajectories are
fixed here from the resonances; not excluding
simplifications in the chosen form for the Odderon.
\section{ Results and discussion}
\subsection {Previous results at $t=0$~\cite{maqm}}
It is our choice to extract from~\cite{maqm}, the useful information
concerning $pp$ and $\bar pp$ at $t=0$. Of course, we do not claim that
better fits are not possible for forward scattering~; many very good
ancient and recent parametrizations, that it is not our scope to discuss
here, are available for $pp$ and $\bar pp$ processes. Thus, as a first
step, we refer to the results found in~\cite{maqm}, where 217 $t=0$ data
(for $pp$ and $\bar pp$) with 4 GeV$\leq\sqrt{s}\leq $1800. GeV
(\cite{data}) have been taken into account, within combined fits of $pp,\,
\bar pp,\, \pi^- p,\, \pi^+p,\, \gamma\gamma,\, \gamma p$ total
cross-sections and $\rho-$ ratios. The selected results under interest
here are given in the Table 1 which exhibits the improvement brought to
the old AQM by the modifications called in the revisited AQM.
\medskip
\newpage
\begin{center}
\begin{tabular}{|l|cccc|}
\hline
Observable &$\sigma^{pp}_{tot}$&$\sigma^{\bar pp}_{tot}$&
$\rho^{pp}$&$\rho^{\bar pp}$\\
Number of points & 85 & 51 & 64 & 17\\
\hline
$\chi^2$ SAQM
&220 &240 &157 &17 \\
$\chi^2$ MAQM
&53 &59 &147 &18 \\
\hline
\end{tabular}
\end {center}
\smallskip

\noindent {\bf Table 1.}
The partial $\chi^2$-s obtained by fitting at $t=0$~\cite{maqm} in
the SAQM (old) and the MAQM (modified).
\bigskip
\noindent
The $\chi^2/d.o.f.$ (for the all processes) was 1.78 in the MAQM (remind
that we fitted our model to the data at $\sqrt{s}\geq 4$ GeV instead of
$\sqrt{s}\geq 5$ GeV in \cite{dglm,cud}).
The recalculated $\chi^2/d.o.f.$
(specifically for $pp$, $\bar pp$ processes) is 1.32 in the MAQM. The
corresponding behavior of $\sigma_{tot}(s)$ and of $\rho(s)=\Re eA(s,0) /
\Im mA(s,0)$ are plotted in~\cite{maqm}. To complete, we give in Table
2 the parameters issued from the combined fits in~\cite{maqm}, used here
for the $pp$ and $\bar pp$ processes.

\medskip
\begin{center}
\noindent
{\small
\begin{tabular}{|l|ccccccc|}
\hline
Parameter & $g_1$(GeV$^{-1}$)  & $g_2$(GeV$^{-1}$)  & $\zeta_P$ &
$g_f$(GeV$^{-1}$)    & $\alpha_f(0)$ &$g_{\omega}$(GeV$^{-1}$)   &
$\alpha_{\omega}(0)$ \\
\hline
Value in & & & & & & & \\
MAQM & 0.3166 & $-$ 0.0239 & 3.396 & 1.112 & 0.8100 & 0.3948 & 0.4217 \\
\hline
\end{tabular}
}
\end{center}
\smallskip

\noindent{\bf Table 2.}
Values of the parameters controlling the MAQM amplitudes at
$t=0$~\cite{maqm}. Recall the Pomeron intercept equals 1. and the values
quoted for $g_f$ and $g_{\omega}$ are coupled to $\lambda_f=0.439$ and
$\lambda_{\omega}=1.$ respectively.
\subsection {MAQM results at $t\neq 0$}
The previous seven parameters are kept fixed to their values determined by
the $t=0$ combined fits. In addition, we fixed from fits to
the resonances those
parameters of the Reggeon trajectories relevant at
$t\neq 0$. Their determination in the present work are recalled in Table 3
(see also Subsection 2.2).
\medskip
\begin{center}
\begin{tabular}{|c||c|c|c|}
\hline
\multicolumn{1}{|c||}{ Reggeon                  } &
\multicolumn{2}{|c| }{"f"                       } &
\multicolumn{1}{|c| }{"$\omega$"                } \\
\hline
\multicolumn{1}{|c||}{ trajectory               } &
\multicolumn{2}{|c| }{square-root   (14)        } &
\multicolumn{1}{|c| }{linear        (16)        } \\
\hline
\multicolumn{1}{|c||}{number of data               } &
\multicolumn{2}{|c| }{3 resonances               } &
\multicolumn{1}{|c| }{2 resonances               } \\
\hline
\hline
parameter&$\gamma_f$&$t_f$&$\alpha'_\omega$\\
         &  5.504 GeV$^{-1}$&14.964 GeV$^{2}$&0.9459 GeV$^{-2}$\\
\hline
\end{tabular}
\end{center}
\smallskip
\noindent {\bf Table 3.} Values of the parameters driving the Reggeon
trajectories at $t\neq 0$, determined by fitting the resonances.
\medskip

Only the remaining parameters are fitted in
isolation from the angular distributions. Of course, the final $\chi^2$
could be improved by refitting the complete set of parameters for the $pp$
and $\bar pp$ processes alone but we decide against doing so.
In this second step, we account for 959 $t\ne 0$ data~\cite{data} with
0.1$<|t|$(GeV$^2$)$\leq 14.2$; 19.4$\leq\sqrt{s}$(GeV)$\leq $630. The
selection $|t|>0.1$ GeV$^2$ is used in order to exclude the
Coulomb-nuclear interference region. It could be included as a
refinement.
After preliminary trials, we select the following conditions for the
current parameterization~:
(i) $\mu=2$ is chosen here because it gives a slightly better $\chi^2$,
but, taking $\mu=1$ does not influence significantly the quality of the
fit: once more the data do not seem precise enough to select a specific
form of the Odderon

(ii) $\nu=1/2$: for this (maximal) value, the Odderon does not violate
unitarity, the results at high energy (LHC) do not seem abnormal and the
Odderon can be considered as an effective phenomenological contribution

(iii) one may fix without damaging the results a unit intercept for the
Odderon ($\delta_{\cal O}=0$) as for our Pomeron and in agreement
with~\cite{Bartels} (iv) linear trajectories are used throughout (except
for the $f$), which represents the more economic (if not the more
efficient) solution

(v) furthermore, as unitarity requires~\cite{unit}, we constrained
$\alpha'_{od}\leq\alpha'_p$.

\noindent
We found the results distributed according to Table 4.

\medskip

\begin{center}
\begin{tabular}{|l|cc|}
\hline
Observable &$({d\sigma/dt})^{pp}$&$({d\sigma/dt})^{\bar pp}$\\
\hline
Number of points & 758 &201 \\
\hline
$\chi^2$  & 1636 & 616 \\
\hline
\end{tabular}
\end{center}

\smallskip
\noindent{\bf Table 4.}
 Resulting partial $\chi^2$-s obtained by fitting the angular
distributions in the MAQM with the parameters in Table~5.
\medskip

\noindent The corresponding $\chi^2/d.o.f.$ (for $t\neq 0$) is 2.38
(for 959 data and 15 parameters listed in Table~5) and
recalculated with the $t=0$ and $t\ne 0$ data together (with the
couplings and intercepts given in Table~2) is 2.19 (for 1176 data and 22
parameters).
\medskip
\begin{center}
\begin{tabular}{|l|l|r|}
\hline
Pomeron& $ b_1$,\  GeV$^{-2}$             &\quad  .1975E+01  \\
& $ b_2$,\  GeV$^{-2}$                    &\quad $-$.2121E+00  \\
&$b_{\zeta_{\cal P}}$,\ GeV$^{-2}$        &\quad  .1227E+01  \\
& $\alpha'_{\cal P}$,\  GeV$^{-2}$        &\quad  .3308E+00  \\
\hline
$f-$Reggeon& $ b_f$,\  GeV$^{-2}$         &\quad  .4094E+01  \\
\hline
$\omega-$Reggeon& $b_\omega$,\
GeV$^{-2}$                                &\quad $\approx $ 0.0\\
 \hline
Odderon& $g_{{\cal O}s}$,\  GeV$^{-2}$    &\quad $-$.1305E$-$02  \\
(soft)& $\zeta_{\cal O}$                  &\quad  .2234E+02  \\
& $\beta_s$,\  GeV$^{-2}$                 &\quad  .5023E+01  \\
& $b_{\cal O}$,\  GeV$^{-2}$              &\quad $-$.1693E+00  \\
& $b_{\zeta_{\cal O}}$,\ GeV$^{-2}$       &\quad  .1417E+01  \\
& $\alpha'_{\cal O}$,\  GeV$^{-2}$        &\quad  .3208E+00  \\
 \hline
Odderon& $g_{{\cal O}h}$,\  GeV$^{-2}$    &\quad  .2434E+01  \\
(hard)& $t_{{\cal O}h}$,\  GeV$^{2}$      &\quad  .4137E+00  \\
& $\beta_h$,\  GeV$^{-2}$                 &\quad  .1197E+01  \\
 \hline
\end{tabular}
\end{center}

\smallskip

\noindent {\bf Table~5.}
Parameters of MAQM obtained in fitting the angular
distributions.

\medskip
\noindent
We found, within the MAQM, a pretty good
reproduction of the data (including the dip and the high $|t|$ regions),
exhibited in Figs.~3-4.
\begin{figure}[ht]\label{fig.3}
\begin{minipage}[t]{7.1cm}
\begin{center}
\includegraphics*[scale=0.38]{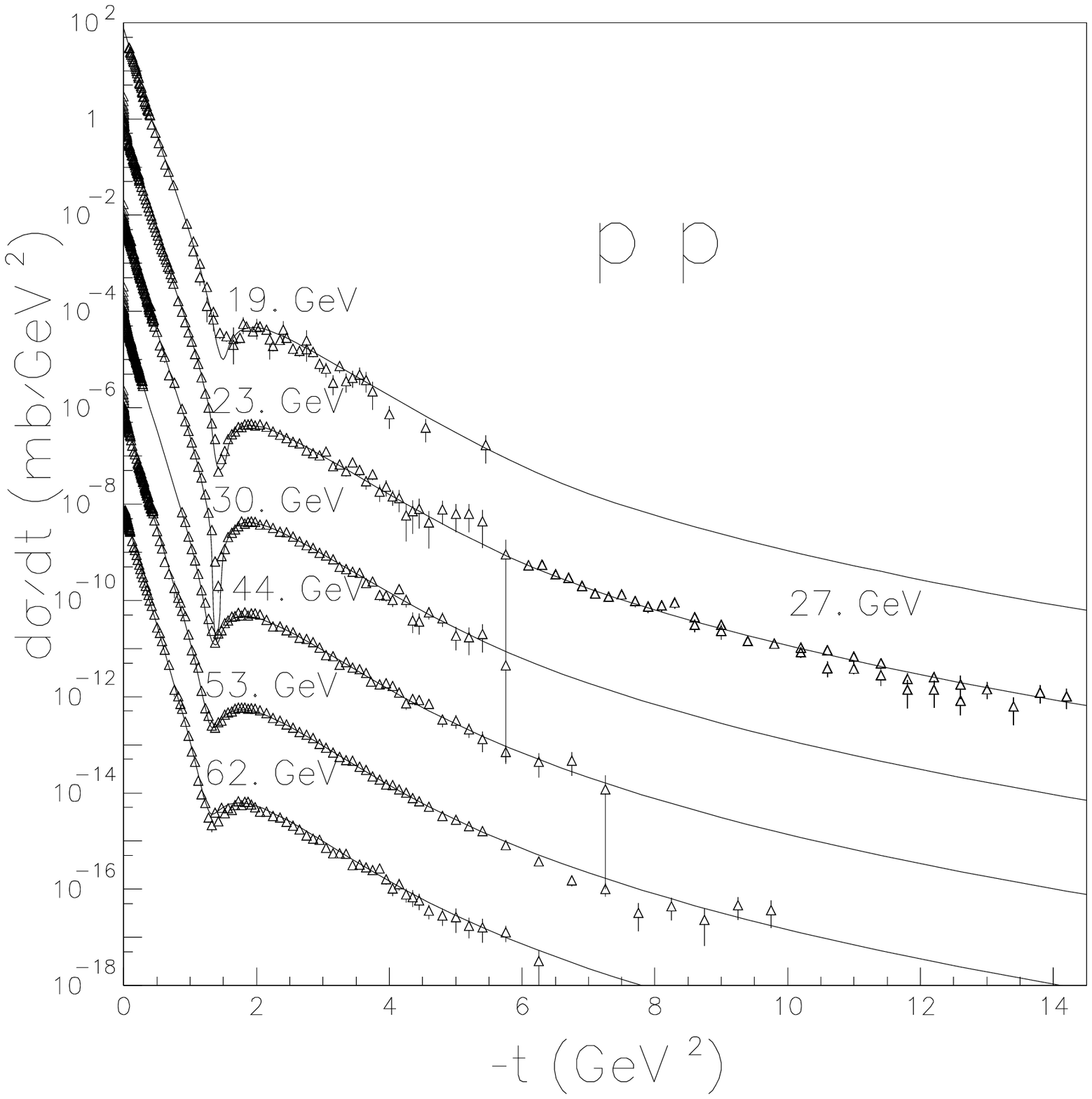}
\caption{Fit of the differential cross-sections of $pp$ interaction,
calculated in MAQM. A factor 10$^{-2}$ between each successive energy
is omitted.}
\end{center}
\end{minipage}
\hskip .8cm
\begin{minipage}[t]{7.1cm}
\begin{center}
\includegraphics*[scale=0.38]{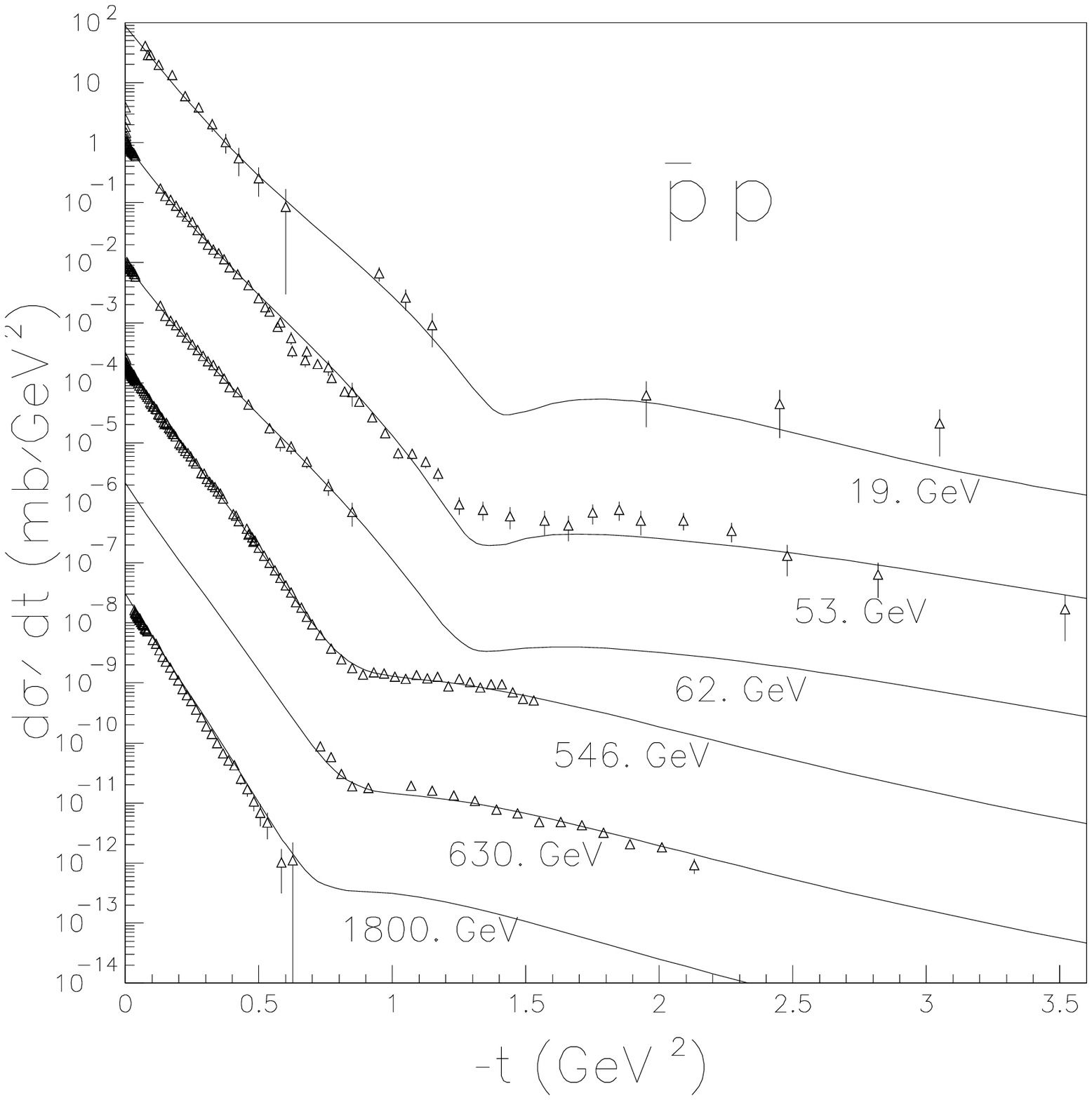}
\caption{The same as in Fig.~3, for $\bar pp$. The Tevatron data are not
fitted.}
\end{center}
\end{minipage}
\end{figure}
\medskip
Searching for an improvement of our results (and accepting to "measure"
the quality of a model by the $\chi^2/d.o.f.$ because we have nothing
better), as already said, it is possible to get a better agreement with
the data when refitting all the parameters together to the forward and non
forward observables, for $pp$ and $\bar pp$ only. In that case, we find
only a non significant improvement in the sense that no modifications is
seen on the figures. Any significant improvement of the $\chi^2$ is not
automatically followed by an improvement visible on the curves. It is of
some interest to compare our results with those of other approaches. A
strict comparison is not easy because, if many of them are available, they
present different objectives, and it is not our aim to discuss the
relative virtues and shortcomings of each work. The model of~\cite{brazil}
for example is an eikonalized model, not operating at the Born level like
ours, in which the energy dependence of the amplitude is absorbed in the
parameters. At the Born level, the nearest models are probably the old
models of~\cite{dl,GaurNic}, based on the Regge theory, furthermore their
results are grossly comparable to ours. The first one contains in
particular a two-Pomeron exchange and other cuts contributions which are
absent in our approach. The second one involves a large number of
parameters and the data at high $|t|$ are not correctly reproduced.
\subsection {Unitarity, Odderon and logarithmic trajectories.}

 -1) As a by-product of the present study, we propose to consider the
scattering amplitude obtained in our MAQM from the point of view
of unitarity and analyticity. To be specific we want to insure that
this amplitude not only respects the Froissart-Martin bound but also
exhibits a zero at low $|t|$ in $\Re eA^{+}(s,t)$, real part of its even
contribution (with respect to the C-parity), as required by a high energy
theorem recently stated by A. Martin~\cite{martin}.

The situation of the first zero of $\Re eA^{+}(s,t)$ is shown in Table 6
for some selected high energies. We agree with the theorem,
with the results quoted in~\cite{martin} and some of
the extensive discussion~\cite{brazil} exhibiting in particular a first zero
of the real part at low $|t|$, decreasing monotonically with the energy.
By also reporting $\Re eA_{\cal P}(s,t)$ in Table 6, we exhibit a
manifestation of the
dominance of the Pomeron at high energies, when the
$f-$Reggeon contribution becomes negligible above the ISR energy range.
\medskip

\begin{center}
\begin{tabular}{|l|c|c|}
\hline
Energy & zero  of $\Re eA^+$ & zero of  $\Re eA_{\cal P}$\\
(GeV) &$-t$  (GeV$^2$) & $-t$ (GeV$^2$)\\
\hline
546& 0.31&0.37 \\
1800&0.27 &0.29 \\
10000&0.21 & 0.22\\
14000&0.21 & 0.22\\
\hline
\end{tabular}
\end{center}

\smallskip
\noindent {\bf Table 6.}
 First values of $-t$ cancelling $\Re eA^{+}(s,t)$ and
$\Re eA_{\cal P}(s,t)$, versus the energy, obtained in the MAQM fit.

\bigskip
 -2) $\Im mA^{+}(s,t)$,
imaginary part of the even component of the amplitude, has an oscillatory
behavior for large $|t|$, in spite of lacking any rescatterings
corrections in the model. This is a consequence of the negative sign of
the new MAQM (comparing with the old SAQM) coupling constant $g_2$ (see
Table~2). Oscillations, or something appearing as diffraction-like
secondary structures are hidden in the differential cross-sections because
of the Odderon contribution dominating in this domain. Of course our
present MAQM model cannot produce any oscillations at large $|t|$, because
it is a model at the Born approximation level, in which the Odderon
dominates at very high energy and in the high $|t|-$region with the
behavior in $\sim 1/t^4$ of the amplitude (see (19)). As a next step, it
would be interesting to eikonalize the model to see in particular if
oscillations in $d\sigma /dt$ appear. We believe that rescattering
corrections (we plan to calculate them in a near future) may be important
at such high energies to be investigated for instance in the TOTEM and
PP2PP projects (see for example~\cite{expfut}) and as a consequence any
extrapolation (in particular those of the angular distributions) may be
doubtful within the present version of non-eikonalized MAQM model.
\bigskip
 -3) As already mentioned above, the "hard" component of the Odderon with
a power behaviour at large $|t|$ looks affectedly from the point of
view of Regge approach (note here that for the whole set of data, the
ratio $|t|/s$ is small, we are in a domain of small angle
scattering, that, we believe, should be described by Regge theory
while the hard Odderon component can be important at larger $|t|$).

\noindent Taking into account the above argument we have considered the
model without its "hard" Odderon component, but for Pomeron and (soft)
Odderon we have tested the nonlinear trajectories with a logarithmic
behavior
\begin{equation}\label{22}
\alpha_i(t)=1+
  \gamma _i\Big[1-(1-t_i/t)\ell n(1-t/t_i)\Big],
  \qquad i={\cal P},\ {\cal O}.
\end{equation}
The minimal value of the threshold $t_{\cal P}$ (there are many
thresholds in $t$-channel for $pp$ and $\bar pp$) should be given by
the $t$-channel physical state with the minimal mass (in our case a
two-pion state, so $t_{\cal P}^{min}=4m_{\pi}^2$). Nevertheless, in
order to take into account the influence of other thresholds we
consider $t_{\cal P}$ as a free (effective) parameter. A similar argument
can be repeated for the Odderon trajectory.

\noindent Logarithmic trajectories mimic a power decreasing amplitude with
$|t|\to \infty$. However, in order to give a sense to such a possibility,
it is necessary to replace the exponential residue functions by power
ones. It leads to an extra number of free parameters. Therefore we used
another, probably oversimplified, method. All exponents, $\exp(b_it)$ in
the Pomeron and in the soft Odderon terms are replaced by
$\exp[(\alpha_{\cal P(O)}(t)-1)b_i]$. Using logarithmic trajectories, we
do not aim to obtain the best fit, rather we only want to check our belief
(and to demonstrate for a reader) that it is then possible to reproduce
large $|t|$ data.

\noindent Resulting $\chi^2/d.o.f=3.39$ as well as agreement with data is
not so bad. The calculated angular distributions are deviated slightly
from the data points mainly around the dips for $pp$ but, as we expected,
are very well reproduced for the large$-|t|$ domain.

\bigskip
\bigskip
Summarizing, we emphasize that the obtained results confirm and reinforce
the conclusion of~\cite{maqm} as a further test of the model: account of
the corrections to a Pomeron-quark interaction and new counting rules for
the secondary Reggeons. In other words using the modified additive quark
model instead of the standard additive quark model, leads to a good
agreement with available experimental data not only at $t=0$, but also at
$t\neq 0$. Besides we found that data on the elastic $pp$ and $\bar pp$
scattering at high energies can be reproduced with a high quality in a
model with the Dipole Pomeron, which has a unit intercept, $\alpha_{\cal P
}(0)=1$ leading to an intermediate growth of the total cross-sections,
$\sigma_{tot}(s)\propto \ell n s$ when $s\to\infty$. Finally, a zero at
small $|t|$ reveals in the real part of the even amplitude, in agreement
with a high energy theorem by Martin.
\medskip

{\large{Acknowledgements.}} We are indebted to E. Predazzi for his support
and discussions. Two of us (MG and EM) would like to thank
the Dipartimento di Fisica Therica of the University of Torino for kind
hospitality. EM thanks also the Institut de Physique Nucl\'eaire de
Lyon for the hospitality. Financial support by the INFN and the MURST
of Italy and from the IN2P3 of France is gratefully  acknowledged.


\end{document}